%% file: paper.tex
\documentclass[%
	reprint,
	superscriptaddress,
	nofootinbib,
	nobibnotes,
	amsmath,amssymb,
	aps,
	pra,
]{revtex4-1}

\usepackage{enumitem}% Include list
\usepackage{graphicx}% Include figure files
\usepackage{dcolumn}% Align table columns on decimal point
\usepackage{bm}% bold math
\usepackage{hyperref}% add hypertext capabilities
\usepackage{acronym}
\usepackage[usenames]{color}

\begin{document}

%%%%%%%%%%%%%%%%%%%%%%%%%%%%%%%%%%%%%%%%%%%%%%%%%%%%%%%%%%%%%%%%%%%%%%%%%%%%%%%%
\title{Improving the background estimation technique in the GstLAL inspiral pipeline with the time-reversed template bank}

%%%%%%%%%%%%%%%%%%%%%%%%%%%%%%%%%%%%%%%%%%%%%%%%%%%%%%%%%%%%%%%%%%%%%%%%%%%%%%%%
\author{Chiwai Chan}
	\affiliation{The University of Tokyo, Hongo 7-3-1 Bunkyo-ku, Tokyo 113-0033, Japan}
	\affiliation{RESCEU, The University of Tokyo, Tokyo, 113-0033, Japan}
\author{Kipp Cannon}
	\affiliation{The University of Tokyo, Hongo 7-3-1 Bunkyo-ku, Tokyo 113-0033, Japan}
	\affiliation{RESCEU, The University of Tokyo, Tokyo, 113-0033, Japan}
\author{Sarah Caudill}
	\affiliation{Nikhef, Science Park, 1098 XG Amsterdam, Netherlands}
\author{Heather Fong}
	\affiliation{The University of Tokyo, Hongo 7-3-1 Bunkyo-ku, Tokyo 113-0033, Japan}
	\affiliation{RESCEU, The University of Tokyo, Tokyo, 113-0033, Japan}
\author{Patrick Godwin}
	\affiliation{Department of Physics, The Pennsylvania State University, University Park, PA 16802, USA}
	\affiliation{Institute for Gravitation and the Cosmos, The Pennsylvania State University, University Park, PA 16802, USA}
\author{Chad Hanna}
	\affiliation{Department of Physics, The Pennsylvania State University, University Park, PA 16802, USA}
	\affiliation{Institute for Gravitation and the Cosmos, The Pennsylvania State University, University Park, PA 16802, USA}
	\affiliation{Department of Astronomy and Astrophysics, The Pennsylvania State University, University Park, PA 16802, USA}
	\affiliation{Institute for CyberScience, The Pennsylvania State University, University Park, PA 16802, USA}
\author{Shasvath Kapadia}
	\affiliation{Leonard E. Parker Center for Gravitation, Cosmology, and Astrophysics, University of Wisconsin-Milwaukee, Milwaukee, WI 53201, USA}
\author{Ryan Magee}
	\affiliation{Department of Physics, The Pennsylvania State University, University Park, PA 16802, USA}
	\affiliation{Institute for Gravitation and the Cosmos, The Pennsylvania State University, University Park, PA 16802, USA}
\author{Duncan Meacher}
	\affiliation{Leonard E. Parker Center for Gravitation, Cosmology, and Astrophysics, University of Wisconsin-Milwaukee, Milwaukee, WI 53201, USA}
\author{Cody Messick}
	\affiliation{Department of Physics, The Pennsylvania State University, University Park, PA 16802, USA}
	\affiliation{Institute for Gravitation and the Cosmos, The Pennsylvania State University, University Park, PA 16802, USA}
\author{Siddharth R. Mohite}
	\affiliation{Leonard E. Parker Center for Gravitation, Cosmology, and Astrophysics, University of Wisconsin-Milwaukee, Milwaukee, WI 53201, USA}
	\affiliation{LSSTC Data Science Fellow}
\author{Soichiro Morisaki}
	\affiliation{Institute for Cosmic Ray Research, University of Tokyo, Kashiwa, Chiba 277-8582, Japan}
\author{Debnandini Mukherjee}
	\affiliation{Leonard E. Parker Center for Gravitation, Cosmology, and Astrophysics, University of Wisconsin-Milwaukee, Milwaukee, WI 53201, USA}
\author{Atsushi Nishizawa}
	\affiliation{The University of Tokyo, Hongo 7-3-1 Bunkyo-ku, Tokyo 113-0033, Japan}
	\affiliation{RESCEU, The University of Tokyo, Tokyo, 113-0033, Japan}
\author{Hiroaki Ohta}
	\affiliation{The University of Tokyo, Hongo 7-3-1 Bunkyo-ku, Tokyo 113-0033, Japan}
	\affiliation{RESCEU, The University of Tokyo, Tokyo, 113-0033, Japan}
\author{Alexander Pace}
	\affiliation{Department of Physics, The Pennsylvania State University, University Park, PA 16802, USA}
\author{Surabhi Sachdev}
	\affiliation{Department of Physics, The Pennsylvania State University, University Park, PA 16802, USA}
	\affiliation{Institute for Gravitation and the Cosmos, The Pennsylvania State University, University Park, PA 16802, USA}
\author{Minori Shikauchi}
	\affiliation{The University of Tokyo, Hongo 7-3-1 Bunkyo-ku, Tokyo 113-0033, Japan}
	\affiliation{RESCEU, The University of Tokyo, Tokyo, 113-0033, Japan}
\author{Leo Singer}
	\affiliation{NASA/Goddard Space Flight Center, Greenbelt, MD 20771, USA}
\author{Leo Tsukada}
	\affiliation{The University of Tokyo, Hongo 7-3-1 Bunkyo-ku, Tokyo 113-0033, Japan}
	\affiliation{RESCEU, The University of Tokyo, Tokyo, 113-0033, Japan}
\author{Daichi Tsuna}
	\affiliation{The University of Tokyo, Hongo 7-3-1 Bunkyo-ku, Tokyo 113-0033, Japan}
	\affiliation{RESCEU, The University of Tokyo, Tokyo, 113-0033, Japan}
\author{Takuya Tsutsui}
	\affiliation{The University of Tokyo, Hongo 7-3-1 Bunkyo-ku, Tokyo 113-0033, Japan}
	\affiliation{RESCEU, The University of Tokyo, Tokyo, 113-0033, Japan}
\author{Koh Ueno}
	\affiliation{The University of Tokyo, Hongo 7-3-1 Bunkyo-ku, Tokyo 113-0033, Japan}
	\affiliation{RESCEU, The University of Tokyo, Tokyo, 113-0033, Japan}

\date{\today}%

%%%%%%%%%%%%%%%%%%%%%%%%%%%%%%%%%%%%%%%%%%%%%%%%%%%%%%%%%%%%%%%%%%%%%%%%%%%%%%%%
\begin{abstract}
Background estimation is important for determining the statistical significance
of a gravitational-wave event. Currently, the background model is constructed
numerically from the strain data using estimation techniques that insulate the
strain data from any potential signals. However, as the observation of
gravitational-wave signals become frequent, the effectiveness of such
insulation will decrease. Contamination occurs when signals leak into the
background model. In this work, we demonstrate an improved background
estimation technique for the searches of gravitational waves (GWs) from binary
neutron star coalescences by time-reversing the modeled GW waveforms. We found
that the new method can robustly avoid signal contamination at a signal rate of
about one per 20 seconds and retain a clean background model in the presence of
signals.
\end{abstract}

\maketitle

\input{acros}

%%%%%%%%%%%%%%%%%%%%%%%%%%%%%%%%%%%%%%%%%%%%%%%%%%%%%%%%%%%%%%%%%%%%%%%%%%%%%%%%
\section{Introduction}
On September 14, 2015, the first detection of gravitational-wave (GW) signal
from the binary black hole (BBH) coalescence~\cite{GW150914} proved that the
BBH mergers occur in the nature and providing us with another way to study the
properties of BHs. Two years later, on August 17, 2017, the GWs and the
accompanied electromagnetic waves from a binary neutron star (BNS)
coalescence~\cite{GW170817, GW170817EM} were also detected for the first time,
marking the start of multi-messenger astronomy informed by GWs. To date, there
are more than 10 GW events due to compact binary mergers were observed in the
first two observing runs~\cite{O2catalog}, and over 50 public alerts of GWs
were issued during the third observing run~\cite{gracedb}, including an event
from a BBH with unequal masses~\cite{GW190412}; these discoveries have
confirmed the possibility of detecting GWs with advanced GW detectors such as
LIGO~\cite{LIGO}, Virgo~\cite{Virgo}, and KAGRA~\cite{KAGRA}. The question of
whether GW exist or not is no longer a concern; instead, the question becomes
how do we detect more GW signals and make more confident detections.

The detection of compact object merger signals is accomplished in part by
perpendicular projection of the data onto the space of waveforms comprising the
family of merger signals of interest; the magnitude of the projection is
referred to as the \ac{SNR}~\cite{maggiore}. When the \ac{SNR} crosses a chosen
threshold, a signal candidate, often called a ``trigger'' is defined and
subjected to further, more computationally costly, scrutiny that ultimately
leads the assignment of a detection ranking statistic.  Since the geometry of
the family of merger waveforms is not well understood, the perpendicular
projection of the data onto their space is approximated by brute-force
projection of the data onto each of a large number of members drawn uniformly
from the space, collectively referred to as a template
bank~\cite{PhysRevD.44.3819,PhysRevD.53.6749,PhysRevD.60.022002,PhysRevD.76.102004,Prix_2007,PhysRevD.80.104014,PhysRevD.89.084041}.

To complete the detection process, we are required to estimate the probability
that the noise produces a GW trigger with a ranking statistic value larger than
or equal to a pre-defined threshold; this probability is known as the
false-alarm probability (FAP) and it describes the statistical significance of
a GW event.  The computation of FAP requires the knowledge of a background
model that describes the statistical properties of noise-induced GW triggers.
If the detector noise is stationary and Gaussian, the FAP of an event can be
computed analytically from the matched-filtering \ac{SNR}.  However, real
detector noise is known to be non-stationary and non-Gaussian over a long
period of time~\cite{Abbott_2016,Abbott_2020}. In this case, we cannot assume
that the \ac{SNR} of the noise triggers are $\chi$-distributed random variables
with two degrees of freedom.

There are various techniques to numerically estimate the background model from
the strain data itself~\cite{pycbc,gstlal_2019}. These techniques try to avoid
picking up any potential signals as noise samples when constructing the
background model. If signals are included in the background model, the search
pipeline will incorrectly believe that the noise process is capable of
producing more false alarms, making the significance estimation more
difficult~\cite{Capano_2017}.  The problem could become worse if the background
model is contaminated by too many signals.  This will be the case for these
techniques as the rate of detectable GW signals increases, and it can be seen
from the analyses described in section \ref{search_results_and_discussion}.

If there existed an alternate space that is orthogonal to the space of merger
waveforms, and that projections onto the alternate space is insensitive to the
presence of genuine signals, but for which the statistical properties of
quantities derived from the projection, such as the \ac{SNR}, remained the same
as for the projection onto the true merger waveform space, then we could
construct the background model from a template bank obtained from that
alternate space and not worry about signal contamination.

It is our conjecture that such spaces exist, and we prove this to be true for
the specific case of BNS signals by explicit construction.  We show that the
use of a time-reversed version of the complete BNS template bank provides an
effective background model that is nearly completely insulated from the
presence of signals in the data.  The statistical properties of the triggers,
such as the \ac{SNR} ($\rho$) and the signal consistency test value ($\xi^2$),
from the outputs of the matched filtering with time-reversed template bank can
be used directly to construct the background model. We illustrate more on the
method in the next section.

This paper is organized as follows. In section \ref{method}, we explain
qualitatively the rationale of using time-reversed template banks to model the
background, and provide proof via an example. In section \ref{tests}, we
describe our experimental setup for the analyses. In section
\ref{search_results_and_discussion}, we demonstrate the robustness of the
improved method in an analysis with an unrealistic signal rate, where about
30000 software simulated BNS signals were injected into a week of strain data.
We also demonstrate an application of our method in another analysis with a
realistic signal rate of one per 1.75 days. Throughout the paper, the data used
in each analysis was real strain data from the second observing run.

%%%%%%%%%%%%%%%%%%%%%%%%%%%%%%%%%%%%%%%%%%%%%%%%%%%%%%%%%%%%%%%%%%%%%%%%%%%%%%%%
\section{Method} \label{method}

\subsection{Time-Reversed Template}
Using the convolution theorem, the matched-filtering in time-domain is a
cross-correlation of the \textit{whitened data} and \textit{whitened
templates}, which is defined as
\begin{equation}
	z_i(t) = \int_{-\infty}^{\infty} \hat{g}^{*}_{i}(\tau)\hat{d}(\tau+t) d\tau
	\label{eq:matched_filter}
\end{equation}
where the hat denotes the whitening process using the single-sided power spectrum
density $S_n(f)$:
\begin{equation}
	\hat{d}(t) = \int_{-\infty}^{\infty} \frac{\tilde{d}(f)}{\sqrt{S_n(|f|)/2}}e^{2\pi i ft} df
	\label{eq:whitening}
\end{equation}
for both strain data $d(t)$ and the $i^{th}$ complex template $g_{i}(t)$; each
component of $g_{i}(t)$, which corresponds to the plus and cross polarizations,
is also normalized to unity in GstLAL. The modulus of Equation
(\ref{eq:matched_filter}) is the \ac{SNR} time series $\rho(t)$ for each
template.

In addition, the ranking statistic used in the GstLAL pipeline also takes a
parameter: $\xi^2$ that we colloquially call it ``chi-squared''. It is designed
to characterize how closely the \ac{SNR} time series resembles the expected SNR
time series computed from the template autocorrelation. It is defined as
\cite{gstlal_2019}
\begin{equation}
	\xi_{i}^{2} = \frac{\int_{-\delta t}^{\delta t}|z_i(t)-R_i(t)|^2 dt}{\int_{-\delta t}^{\delta t}(2-2|R_i(t)|^2)dt}
	\label{eq:xi2}
\end{equation}
where $\delta t$ is a tunable time window and $R_{i}(t)$ is the template
autocorrelation function defined by
\begin{equation}
	R_{i}(t) = \int_{-\infty}^{\infty} \frac{|\tilde{g}_{i}(f)|^2}{S_n(|f|)} e^{2\pi i ft} df.
\end{equation}
The $\rho$ and $\xi^2$ are mostly the quantities that are used to numerically
construct the background model. Hence, it is our goal to construct an estimate
of the background statistic of $\rho$ and $\xi^2$.

We propose using the time-reversed version of original search template bank to
construct the background model for BNS searches because of the following
reasons. First, the inspiral-merger stage of a low mass binary such as BNS is
typically of the order of hundreds of seconds, and the chirp waveform is not
symmetric in time. The inner product of a original (forward) template and its
time-reversed copy will be small. Matched-filtering with a time-reversed
template will cause the peaks in the \ac{SNR} time series produced by the
signals to be strongly suppressed. Second, the features in the original
waveform such as the amplitude and duration are not lost. The response of
matched filter to the Gaussian noise and glitches (at least for a
time-symmetric glitch as we have shown in the following example) should remain
similar.

\subsection{Responses to Signals}
\begin{figure}
	\centering
	\includegraphics[width=\linewidth]{{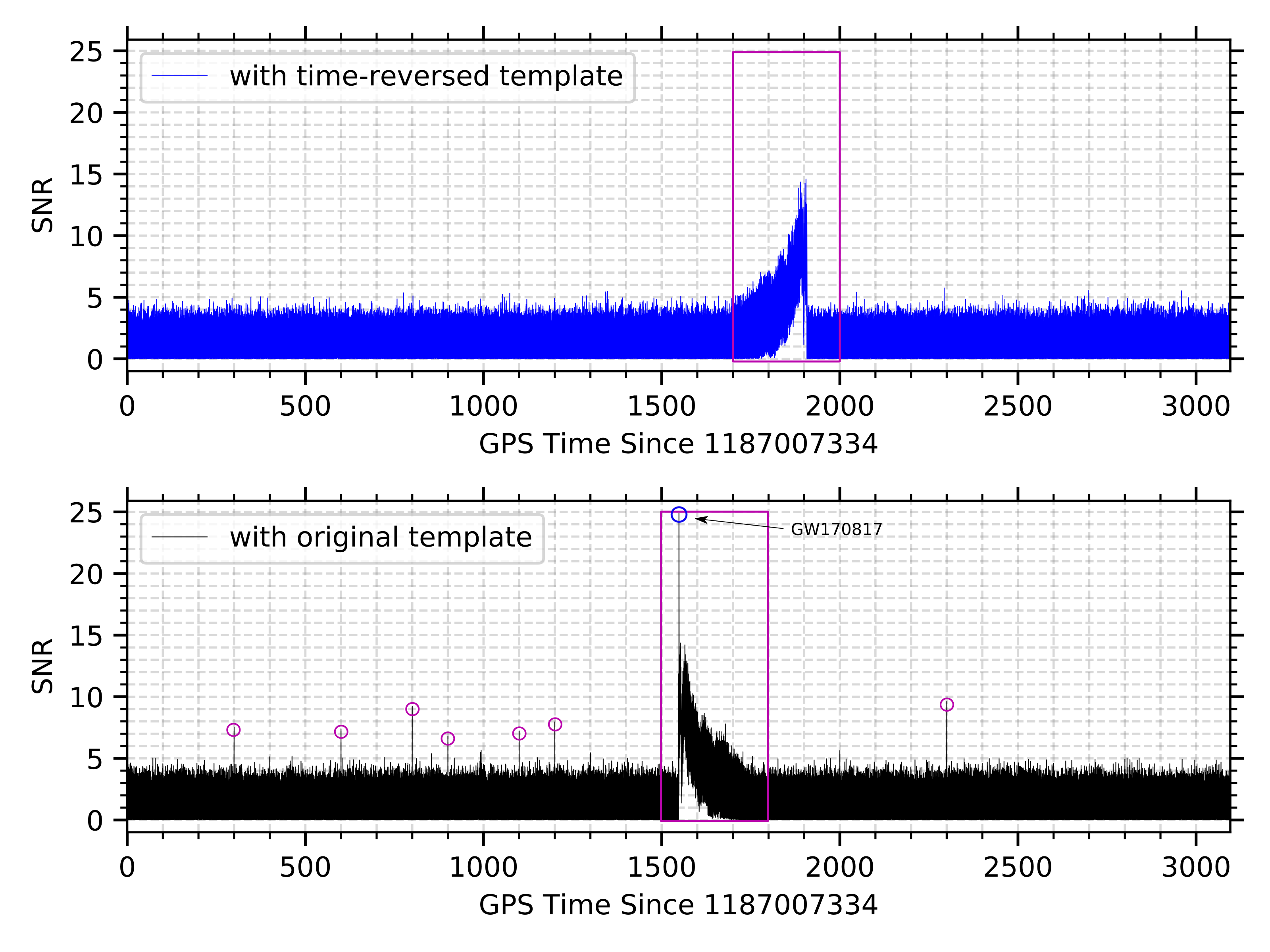}}
	\caption{The matched-filter output using the time-reversed template (upper
	row) and the original template (bottom row). The ``peaks'' marked by the
	pink circles represent the injected signals that are identified with
	visually prominent \ac{SNR}; the blue circle represents the event GW170817.
	The region marked by the pink boxes is the matched-filtering response of a
	time-symmetric glitch. It persists in the matched-filitering outputs of both
	original and time-reversed template, but occurs in different time and is
	time-reversed.}
	\label{fig:response_to_signals}
\end{figure}

As a demonstration that the time-reversed template is not sensitive to signals,
we inject simulated GW signals from BNS, generated from \texttt{TaylorF2}
waveform family \cite{TaylorF2}, into the strain data every 100 seconds, and
perform matched-filtering with a template that has the same parameters as the
injections. The strain data is a segment of real data from LIGO Livingston
detector around the event GW170817~\cite{gwosc}, which was known to contain
a glitch. The injected GW signals are spinless BNS located uniformly from 90
Mpc to 100 Mpc and have the same masses of about $1.78M_\odot$ and
$1.08M_\odot$; the range of distance is chosen that the BNS can produce
visually recognizable peaks, and the value of the masses is unimportant as long
as they represent the mass of a typical BNS, but we chose them to be the same
masses as the event GW170817.  The output of the matched-filtering using the
time-reversed template and the original template is shown in Figure
\ref{fig:response_to_signals}.

The result shows that the injections and the real signal produce visually
prominent peaks (circles in Figure \ref{fig:response_to_signals}) in the
\ac{SNR} time series when filtering with the original template that matches the
injections and real signal, but none of those peaks are identified in the case
of time-reversed template. Moreover, the glitch, which is highlighted by a pink
box in Figure \ref{fig:response_to_signals}, still persists in the output of
matched-filtering using time-reversed template, but occurs at a different time
and it is time-reversed as well; this indicates that the glitch is
time-symmetric. However, the time of occurrence of the glitch is irrelevant; we
are only concerned about the matched-filitering statistic, and a time-reversed
matched-filtering output for the glitch does not affect the statistic.
Therefore, the result suggests that the time-reversed version of the original
template is insensitive to the injections and the real signal even though the
original template has parameters that match the signal, and the statistics of a
time-symmetric glitch can be preserved.

\subsection{Responses to Noise}
\begin{figure}
	\centering
	\includegraphics[width=\linewidth]{{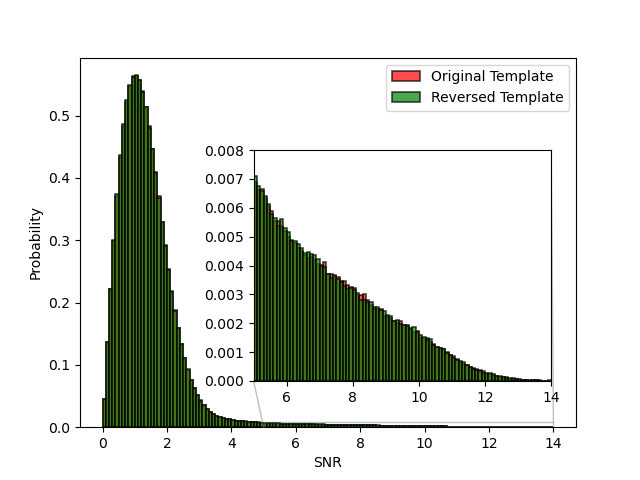}}
	\caption{The histogram of \acp{SNR} with a uniform binning of size 0.1 for both
	original template and time-reversed template. The \ac{SNR} produced by the
	time-reversed template agrees well with the one produced by the original
	template.}
	\label{fig:histogram}
\end{figure}

To show the matched-filter's response against noise, we histogrammed the
previous result and normalized it by the total counts. Figure
\ref{fig:histogram} shows the statistical distribution of \acp{SNR} produced by
the original and time-reversed templates. Although the data is injected with
simulated signals, the \acp{SNR} contributed by the simulations are the
minorities when comparing to the \acp{SNR} produced by the noise component of
the data. Thus, the \ac{SNR} contributed by the simulations is small compared
to the \ac{SNR} contributed by noise.

From the histogram, we note the distribution of \acp{SNR} of the original and
time-reversed template agrees well with each other. This ensures that the
time-reversed templates preserve the \ac{SNR} distribution for noise, and it
should not negatively impact the evaluation of likelihood ratio and FAP
and/or FAR~\footnote{If we assume the triggers produced by noise follows a
Poisson distribution, the FAP can be mapped to the \emph{false-alarm rate}
(FAR) which describes the mean rate at which the noise produces at least M
triggers with ranking statistic value larger than or equal to a threshold.}.

%%%%%%%%%%%%%%%%%%%%%%%%%%%%%%%%%%%%%%%%%%%%%%%%%%%%%%%%%%%%%%%%%%%%%%%%%%%%%%%%
\section{Tests} \label{tests}

\subsection{Experimental Setup}
The matched-filter's responses to signals and noise for the time-reversed
template have presented evidence that it is insensitive to the GW signals and
able to remain the noise statistic for one template. We would like extend it to
a complete GstLAL inspiral offline search and examine the performance of
modeling background using a time-reversed template bank.

The background collection infrastructure was slightly modified to allow the
collection of background during the times when only one detector is online
(single-detector times). The modification is intended to show that the current
technique should not collect background samples during single-detector times as
it is impossible to form coincidences with single detector. Nevertheless, the
improved technique using a time-reversed template bank should possess the
ability to use the data during single-detector times even if signals are
present.

\subsubsection{Template Bank} \label{template_bank}
Banks of search templates are generated from the \texttt{TaylorF2}
post-Newtonian waveform family~\cite{TaylorF2} and
\texttt{SEOBNRv4\_ROM}~\cite{SEOBNRv4_ROM}. For binaries with chirp mass $0.0
M_\odot \leq \mathcal{M} \leq 1.73 M_\odot$, the templates are generated with
\texttt{TaylorF2}. For binaries with chirp mass $1.73  M_\odot <
\mathcal{M} \leq 1000.0 M_\odot$, the templates are instead generated with
\texttt{SEOBNRv4\_ROM}. The template bank is comprised of 69781 templates
covering a wide range of masses, including the BNS range. The time-reversed
template bank is constructed from the same search template bank.

\begin{table*}
\begin{ruledtabular}
	\centering
	\begin{tabular}{ccccccc}
		Mass 1 ($\text{M}_\odot$) & Mass 2 ($\text{M}_\odot$) & Spin 1 & Spin 2 & Detector & GPS Time & Distance (Mpc)\\
		\hline
		$1.4096$ & $1.4036$ & $0$ & $0$ & H1 & 1176297508 & 170.6\\
		$1.4031$ & $1.4078$ & $0$ & $0$ & H1 & 1176457885 & 26.9\\
		$1.3987$ & $1.4063$ & $0$ & $0$ & L1 & 1176587931 & 103.4\\
		$1.3978$ & $1.4127$ & $0$ & $0$ & H1 & 1176739231 & 54.2\\
	\end{tabular}
	\caption{The four BNS injections in the injection set B. The first fours
	columns are the masses and spins of the injections, and the last two columns
	are the times where the injections were added and the distances from Earth. The
	column ``Detector'' represents which detector was on for detection during the
	injection time.}
	\label{table:injection_b}
\end{ruledtabular}
\end{table*}

\subsubsection{Gravitational Wave Data}
We use a week of strain data from Hanford (H1) and Livingston (L1) in the
second observing run, beginning on April 14 21:25:00 GMT 2017 and ending on
April 21 21:25:00 GMT 2017~\cite{gwosc}. This chunk of data is found to contain
no significant events by the GstLAL inspiral pipeline using the search template
bank mentioned in \ref{template_bank}, so a collection of software simulated GW
signals from BNS will be added to the data to mimic the presence of actual GW
signals; these injections are generated from the
\texttt{TaylorT4ThreePointFivePN} waveform model. There are two injection sets
prepared for the analysis, each serving a different purpose.

\subsubsection{Injection Set A}
Injection set A contains 30240 spinless binary sources with typical neutron
star (about 1.4 $\text{M}_{\odot}$) located uniformly in distance and ranged
from 20 Mpc to 200 Mpc. The masses of these injections were generated from a
Gaussian distribution with a mean of 1.4 $\text{M}_{\odot}$ and a standard
deviation of 0.1 $\text{M}_{\odot}$. Injections were added every 20s without
considering detector duty cycles. Due to detector downtime, only 20592
injections occur at times the detectors were observing. The amount of
injections over a week of data is unrealistic, but it can serve as a test for
the robustness of the background estimation technique with time-reversed
template bank (henceforth referred to as the improved technique).

\subsubsection{Injection Set B}
Injection set B contains only 4 spinless BNS sources having the parameters
listed in Table \ref{table:injection_b}. These injections were specifically
added to the strain data during single-detectors times. The improved technique
should be able to use the triggers during the single-detector times to model
the background without contaminating it with signal-like triggers.  Therefore,
injection set B serves as a test to determine whether or not the improved
technique can achieve that goal. The rate of the GW signals in this test (4 BNS
injections in a week of data) is realistic when considering the mean rate of
events detected by GstLAL during the third observing run. So the test can be
used to demonstrate the usefulness of the improved technique in a practical
situation.

\subsubsection{Searches}
For the following searches, we define the \emph{original technique} as the background
estimation technique with original templates and the \emph{improved technique} as the
background estimation technique with time-reversed templates.

%%%%%%%%%%%%%%%%%%%%%%%%%%%%%%%%%%%%%%%%%%%%%%%%%%%%%%%%%%%%%%%%%%%%%%%%%%%%%%%%
\section{Search Results and Discussion} \label{search_results_and_discussion}

\subsection{The Search with Severe Signal Contamination}
To test the robustness of the improved technique, injection set A was added to
the original data for the analysis. In this subsection, we will first show the
result of the search using the original technique. Then, the result of the same
search using the improved technique will be shown for comparison.

\subsubsection{Searches with Original Technique}
As a reference, we first conducted the search using the original technique. The
result of the search is summarized in the plot of the event counts versus
$\text{ln}\mathcal{L}$ threshold (Figure \ref{fig:rate_vs_threshold_injected} :
top). From the figure, we note that the curve of the observed event count
starts to deviate from the noise model prediction at around
$\text{ln}\mathcal{L} = -4$, suggesting that there are more events observed
with a $\text{ln}\mathcal{L}$ above that threshold than the prediction by the
noise model. Since the observed events could be thought of as the sum of signal
and noise events, the extra events will be the signal-like events. To properly
define the meaning of a true GW event, we can require that all true GW events
to have a FAR $\leq 3.858\times 10^{-7}$ (1 false alarm per 30 days). With this
threshold, we found that 2191 events were below the threshold, meaning that
only about $10.6\%$ injections were recovered by the original search pipeline.

\subsubsection{Searches with Improved Technique}
\begin{figure}
	\includegraphics[width=1.0\linewidth]{{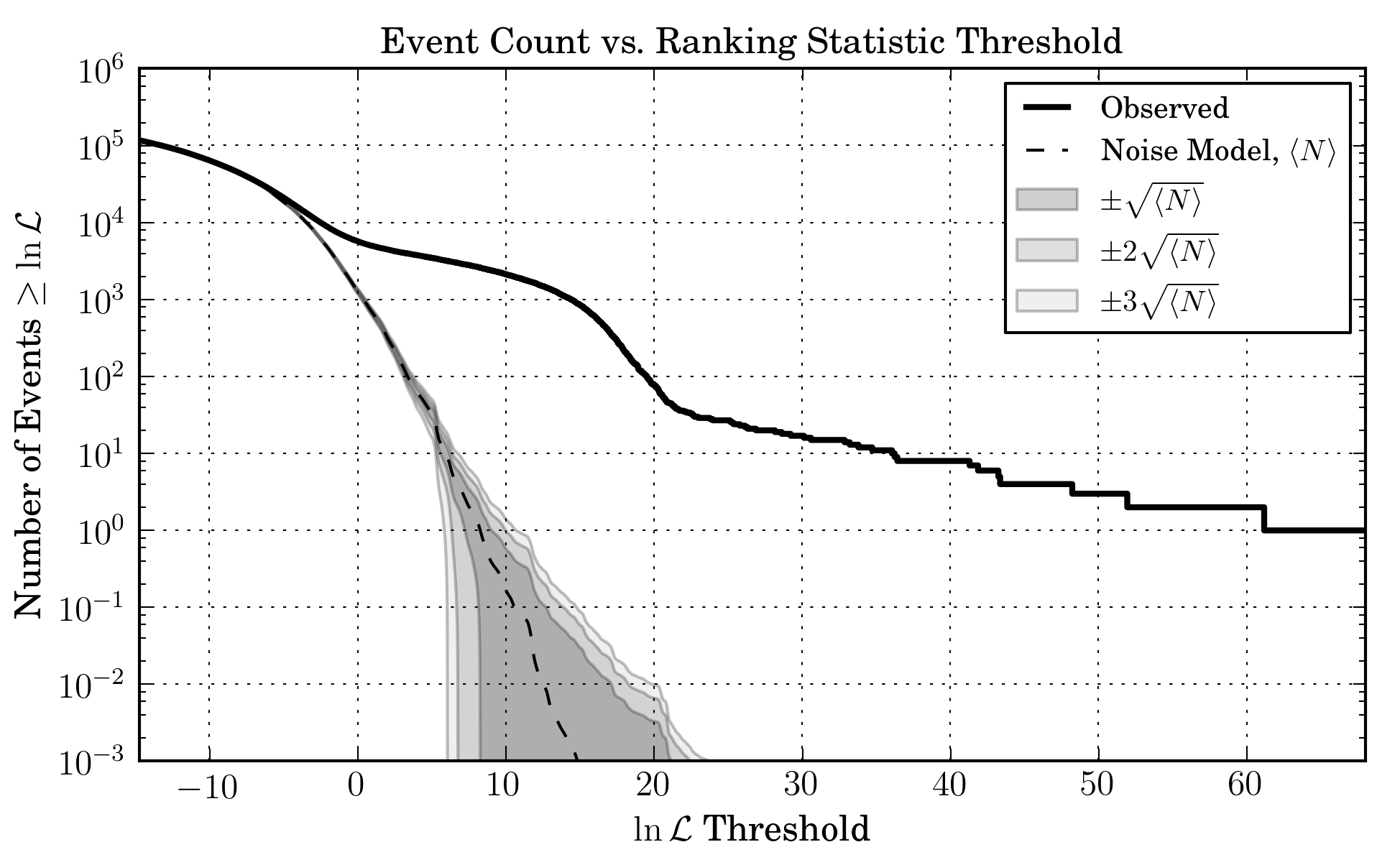}}
	\includegraphics[width=1.0\linewidth]{{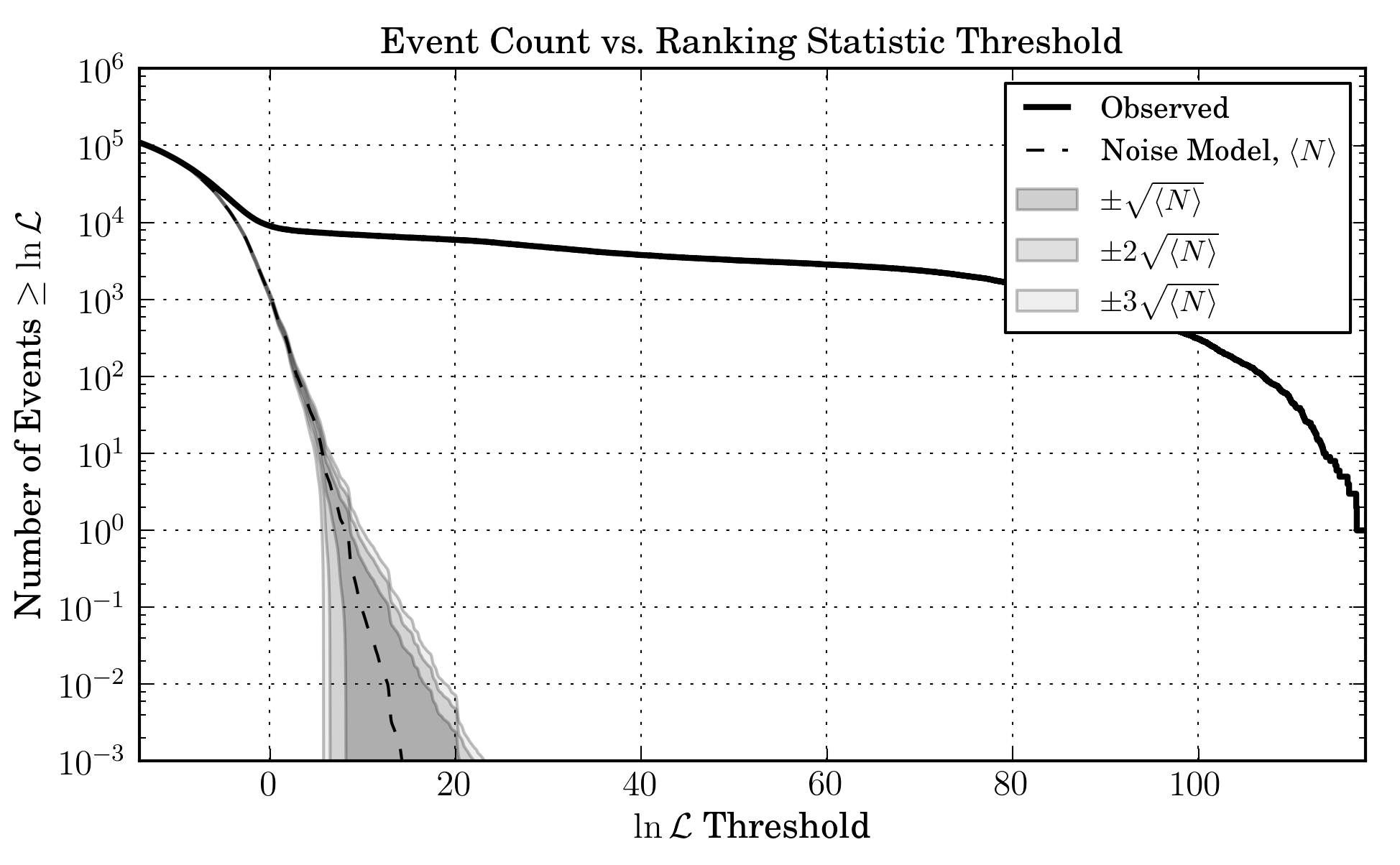}}
	\caption{The plots of the event count vs. ranking statistic threshold for the
	injection analysis using the \textit{original technique} (top) and
	\textit{improved technique} (bottom). There are more injections identified with
	the improved technique compared to the original technique.}
	\label{fig:rate_vs_threshold_injected}
\end{figure}

Next, we performed the search again but using the improved technique. The event
count versus $\text{ln}\mathcal{L}$ threshold plot (Figure
\ref{fig:rate_vs_threshold_injected} : bottom) shows more injections were
recovered and with higher $\text{ln}\mathcal{L}$. If we set the same FAR
threshold for the events to be true GW events, then there were 6990 events
below the threshold, which means $33.9\%$ injections are recovered. Therefore,
the improved technique is able to find more injections than the original
technique.

\begin{figure}
	\includegraphics[width=\linewidth]{{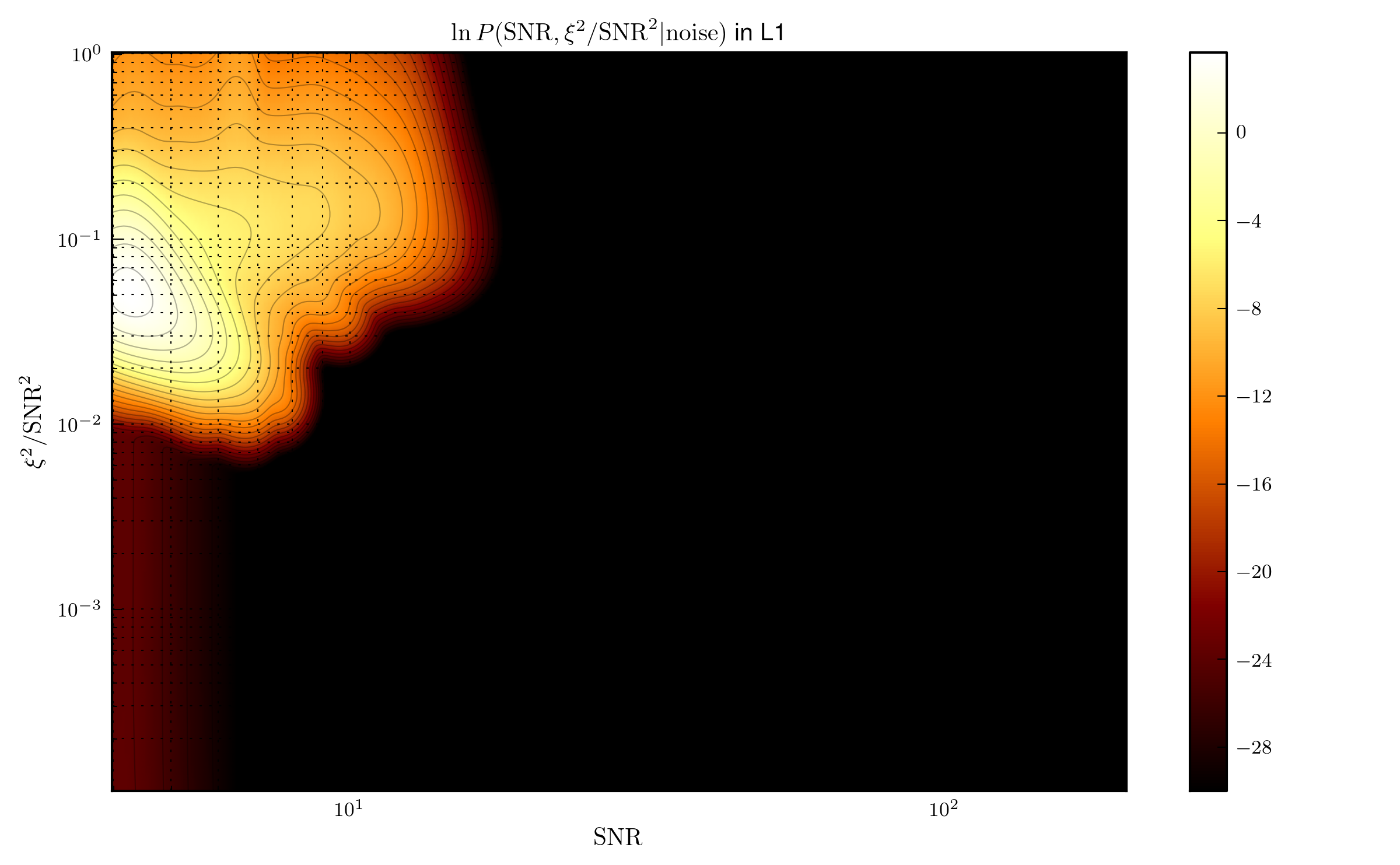}}
	\includegraphics[width=\linewidth]{{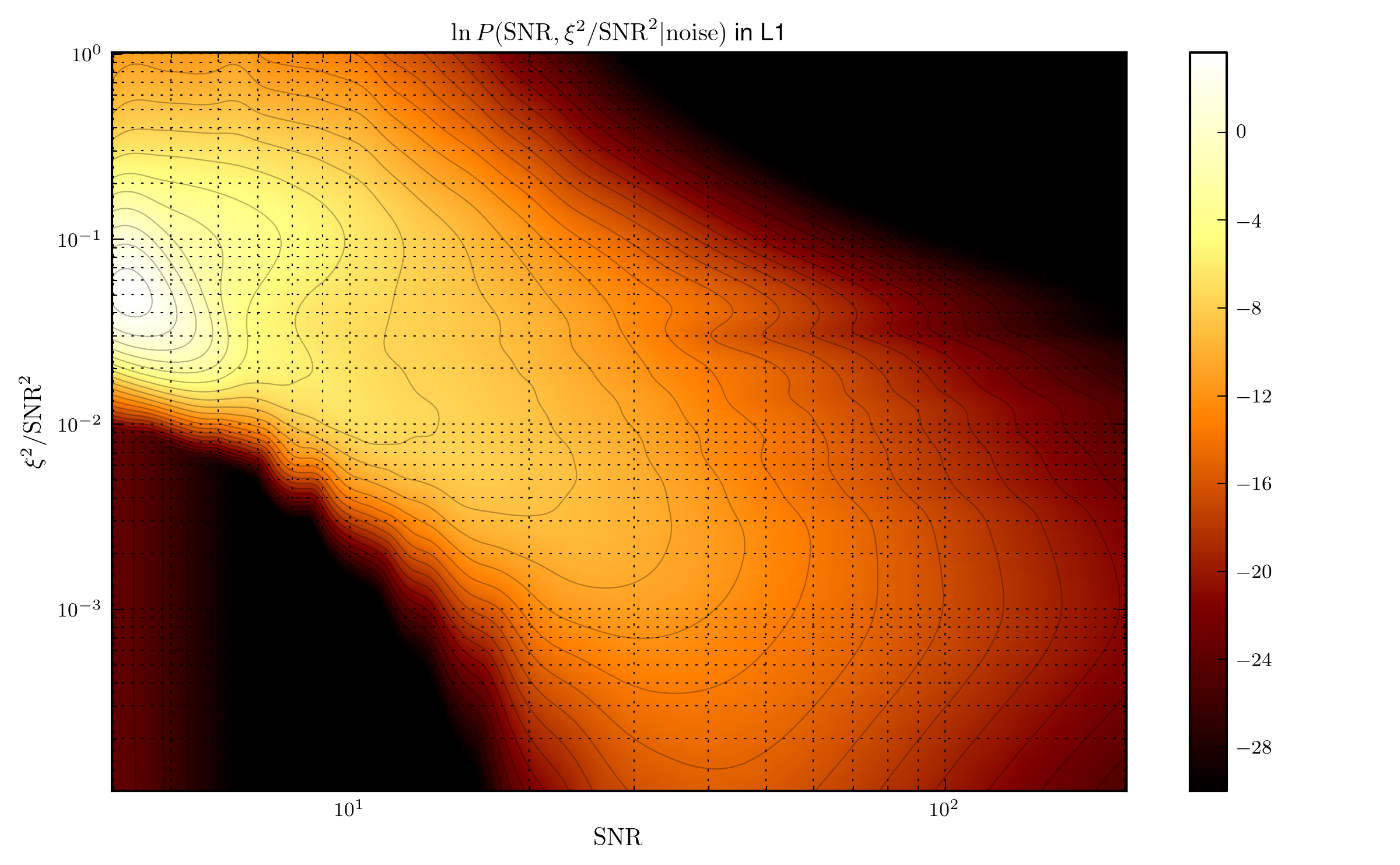}}
	\includegraphics[width=\linewidth]{{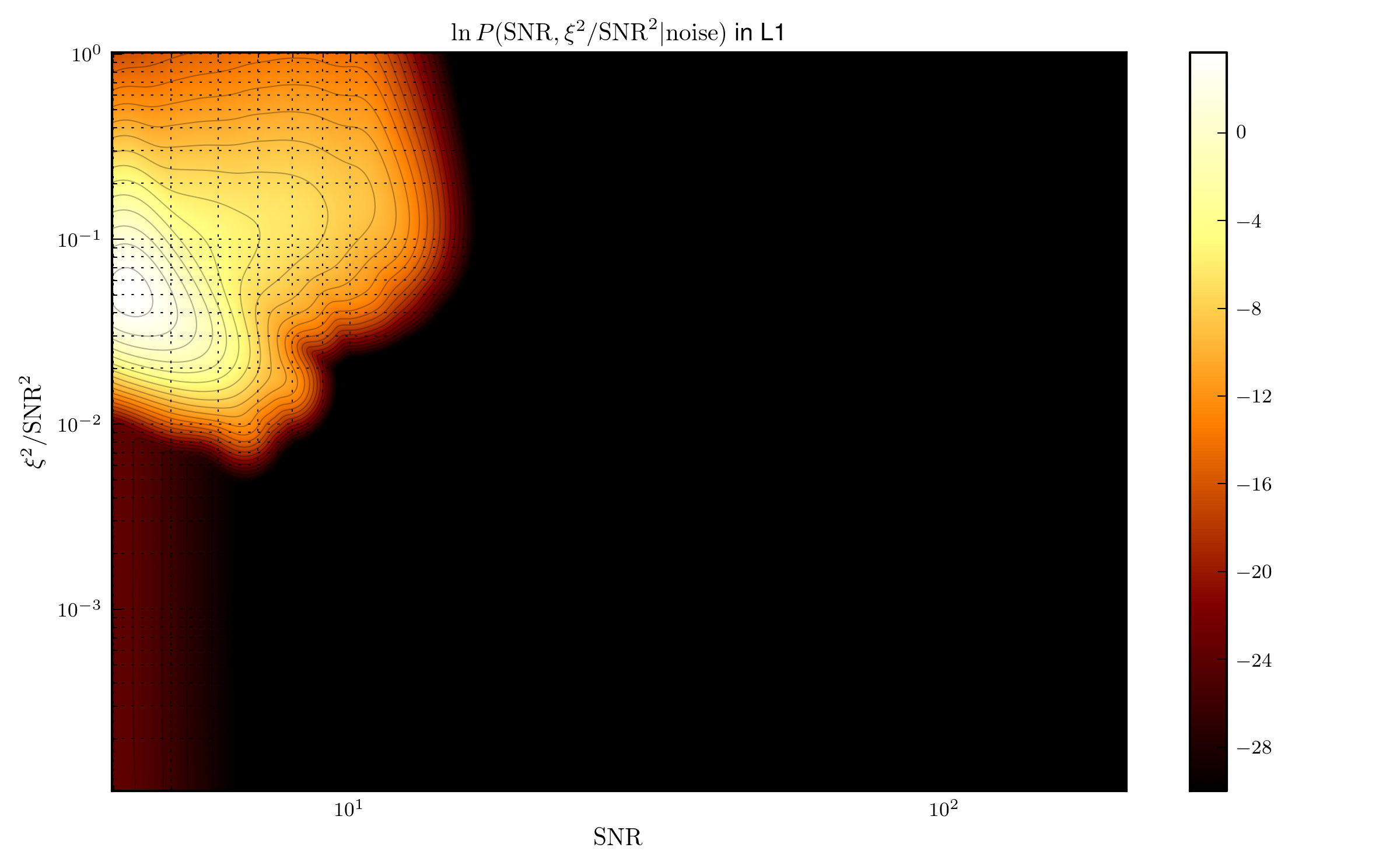}}
	\caption{The background PDFs for the same template bank bin using different
	techniques. Top: the cleanest background PDF we could obtain from the GstLAL
	inspiral pipeline using original technique since there no injections in the
	data when estimating the PDF. Middle: the background PDFs obtained from the
	data with injections using the original technique. Bottom: the background
	PDFs obtained from the data with injections using improved technique. There
	is an extended island, which is characteristic of signal-like triggers,
	filling up the middle figure, but it is not found on the top and bottom
	figures. This suggests that the characteristic of signal-like triggers is
	removed by the improved technique and the ``true'' background is sufficiently
	restored.}
	\label{fig:PDFs}
\end{figure}

\subsubsection{Discussion}
To understand why the technique using the time-reversed template banks is able
to improve the the search result, we can examine the differences in the
background PDFs between the two techniques. Figure \ref{fig:PDFs} shows three
background PDFs for the same region of the parameter space using different
techniques. We note that the original data contains no GW signals, so the
background PDFs obtained from the original data before adding injections is the
cleanest background we could obtain from the GstLAL inspiral pipeline using the
original technique. We will refer this background as the \textit{optimal
background} and it is shown on the top of Figure \ref{fig:PDFs}. The background
PDFs estimated from the data with injections using the original and improved
techniques are shown on the middle of Figure \ref{fig:PDFs} and the bottom of
Figure \ref{fig:PDFs}; they are refered to the \textit{contaminated background}
and the \textit{recovered background} respectively.

From the contaminated background and optimal background, we can see that the
background PDF obtained using the original technique is contaminated by the
injections. Any true GW trigger with ($\rho$, $\xi^2/\rho^2$) value falls into
the contaminated region will be penalized by the likelihood-ratio ranking
statistic, and tends to be classified as a noise-trigger. The consequence of
the contamination is the decrease of the number of events at a given
$\text{ln}\mathcal{L}$ threshold (compare Figure
\ref{fig:rate_vs_threshold_injected} : top and Figure
\ref{fig:rate_vs_threshold_injected} : bottom). On the other hand, the
recovered background resembles the optimal background but different from the
contaminated background. This suggests two results: the improved technique
using the time-reversed template bank can estimate the ``true'' background even
though the data is full of injections, and the recovered background can be
similar to the optimal background.

\begin{figure}
	\centering
	\includegraphics[width=\linewidth]{{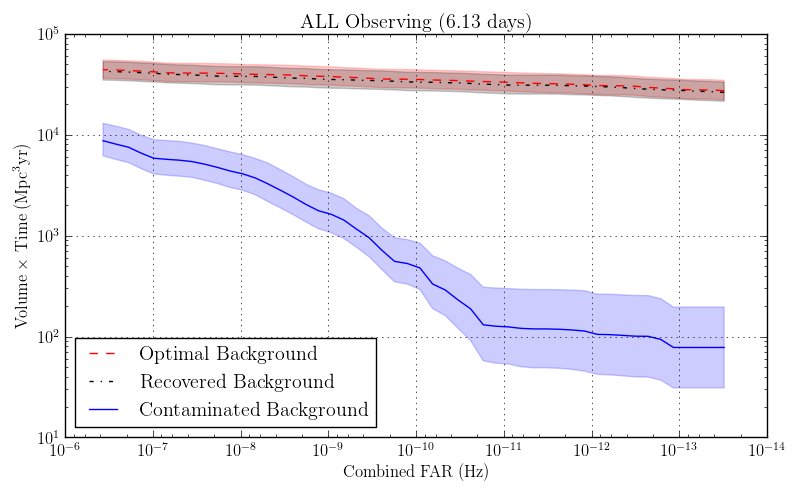}}
	\caption{The plot of the sensitive volume-time versus the combined FAR for
	the three different backgrounds. The sensitivity of the pipeline using the
	improved technique (with recovered background) is uniformly better than the
	pipeline using the original technique (with contaminated background). The
	improved technique is able to achieve the optimal sensitivity in the presence
	of injections.}
	\label{fig:20000_VT}
\end{figure}

The plot of the sensitive volume-time versus the combined FAR (Figure
\ref{fig:20000_VT}) gives a comprehensive comparison of the sensitivity of the
pipeline among the three different backgrounds at different FAR thresholds.
From the sensitivity plot, we see that the analysis using recovered background
is uniformly more sensitive than the contaminated background. This implies that
the sensitivity of the search pipeline is greatly improved as a result of this
improved technique. On the other hand, the optimal sensitivity is consistent with
the sensitivity using the improved technique, sugguesting that the improved
technique is able to achieve the optimal sensitivity even in the presence of
many injections.

\subsection{The Search with Realistic Signals Contamination}
The previous results have demonstrated the robustness of the improved
background estimation technique in the presence of many signals, we will focus
on the performance in the realistic signal contamination in this subsection. In
this search, the injection set B (Table \ref{table:injection_b}), which
contains three BNS injections in H1 and one BNS injection in L1, were added to
the data during the single-detector times. We will continue referring the
backgrounds obtained from the original and improved techniques as the
contaminated background and recovered background respectively, and the cleanest
background that can be obtained by the pipeline is referred to optimal
background. They will be used to understand the following search results.

\subsubsection{Search with Optimal Background}
Using the optimal background mentioned previously, the search result reveals
that Hanford (H1) identified an event occurred roughly at GPS time 1176457885
with $\text{ln}\mathcal{L} = 21.7$ and FAR $=4.01\times10^{-7}$ per second; it
is an event that is signficant enough to be considered as recovered.  The
identification time of this event coincides with one of our injections in H1,
which suggests that the pipeline can only recover at most one injection even
with the optimal background. The remaining three injections were not identified
even in the optimal background because there were not triggers around the
injection times, which implies that those injected signals might not be strong
enough to be detected or no templates with parameters could match the
injections.

\begin{figure}
	\includegraphics[width=\linewidth]{{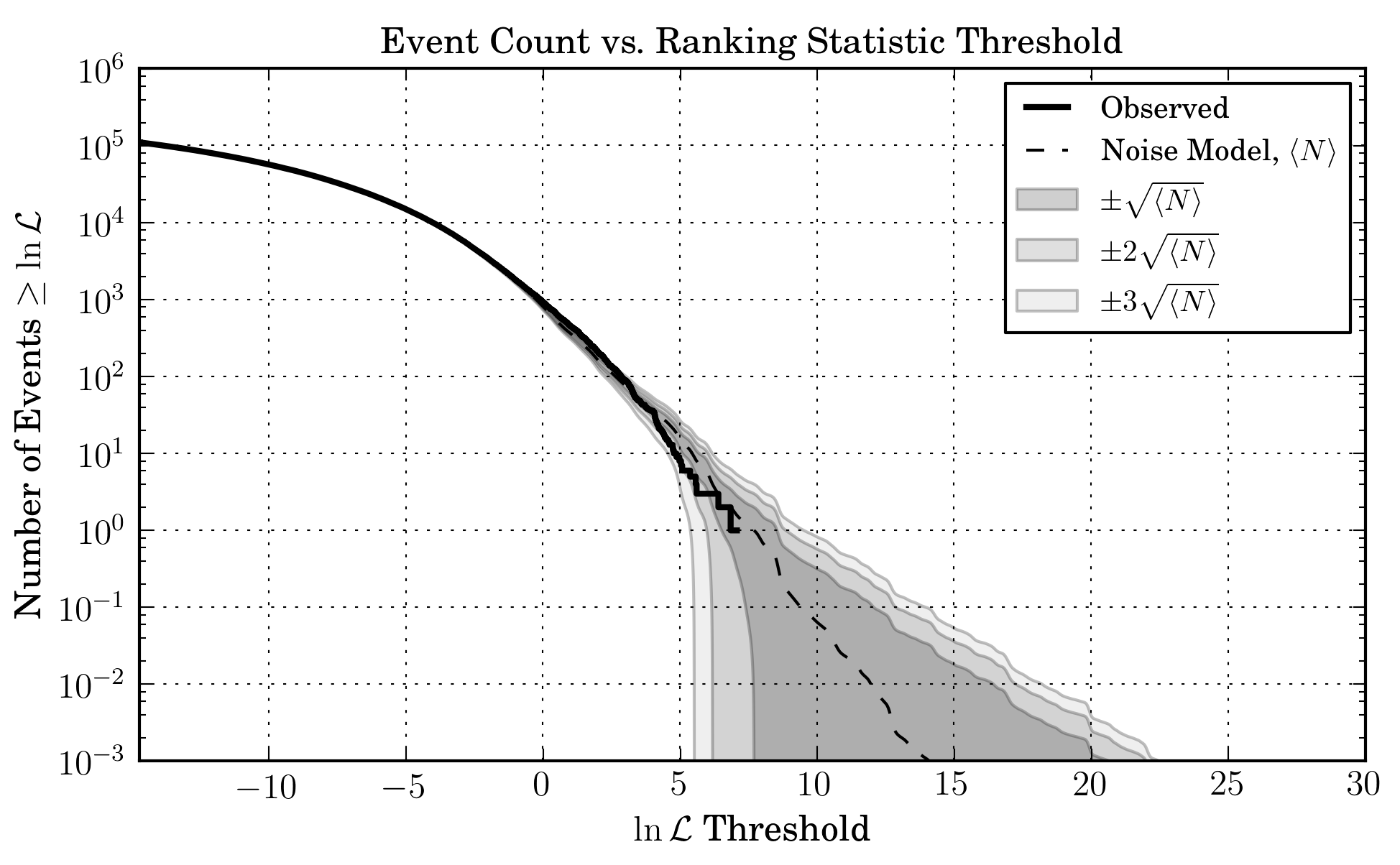}}
	\includegraphics[width=\linewidth]{{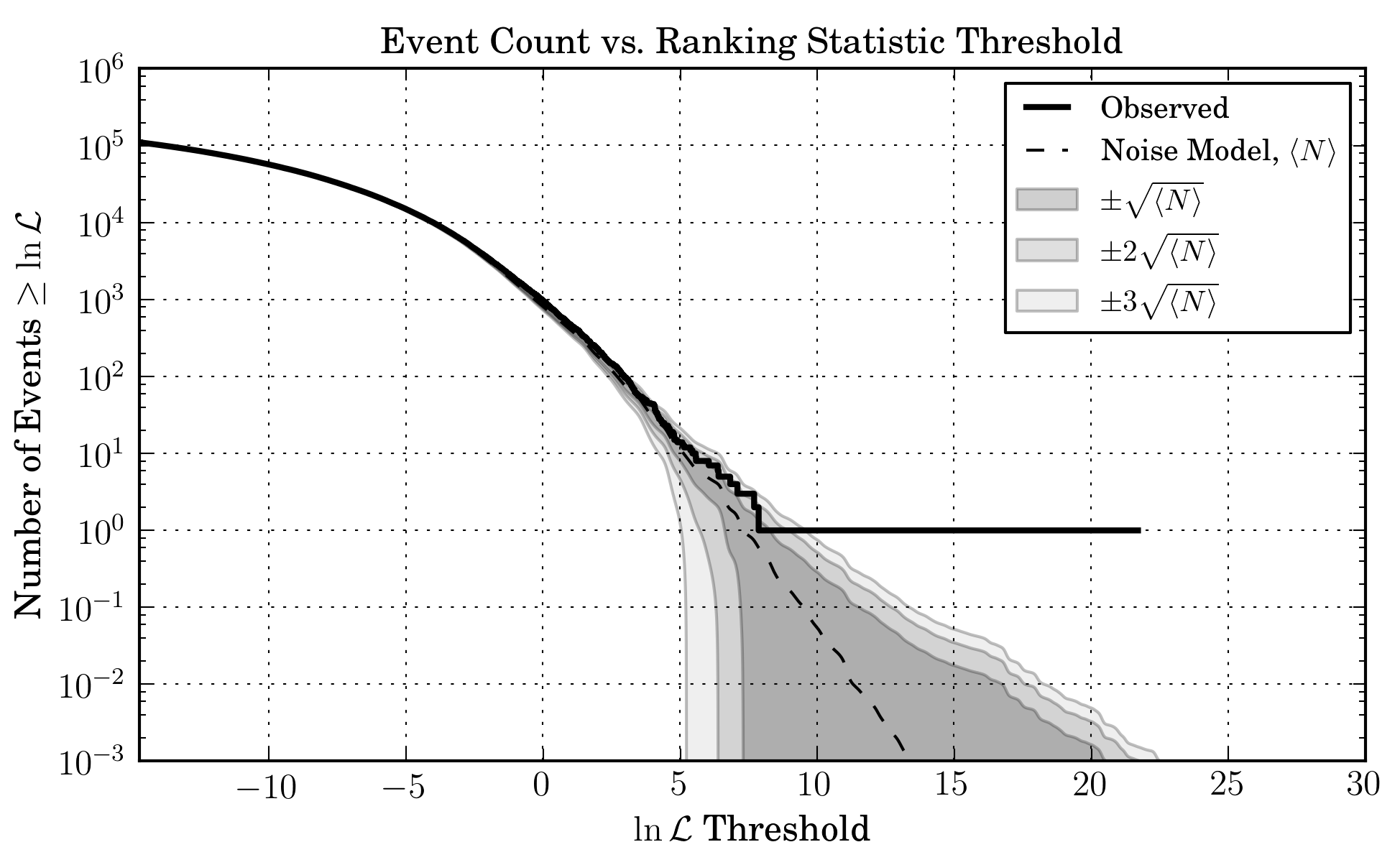}}
	\caption{The plots of the event count vs. ranking statistic threshold for the
	same analysis but using the original technqiue (top) and the improved
	technique (bottom). There was an injection identified with the improved
	technique, while there are no injections identified with the original
	technique.}
	\label{fig:signal_contamination}
\end{figure}

\subsubsection{Search with Original Background}
The search using the contaminated background was not able to identify
any event (Figure \ref{fig:signal_contamination} : top). In particular, the
same event is not identified as a significant trigger anymore: it now has
$\text{ln}\mathcal{L} = -3.49$ and FAR $=1.00$ per second.

\subsubsection{Search with Improved Background}
On the other hand, the search using the recovered background was able to
identify the same event with high significance (Figure
\ref{fig:signal_contamination} : bottom). The event is now found with
$\text{ln}\mathcal{L} = 21.7$ and FAR $=2.62\times10^{-7}$ per second.

\subsubsection{Discussion}
\begin{figure}
	\includegraphics[width=1.0\linewidth]{{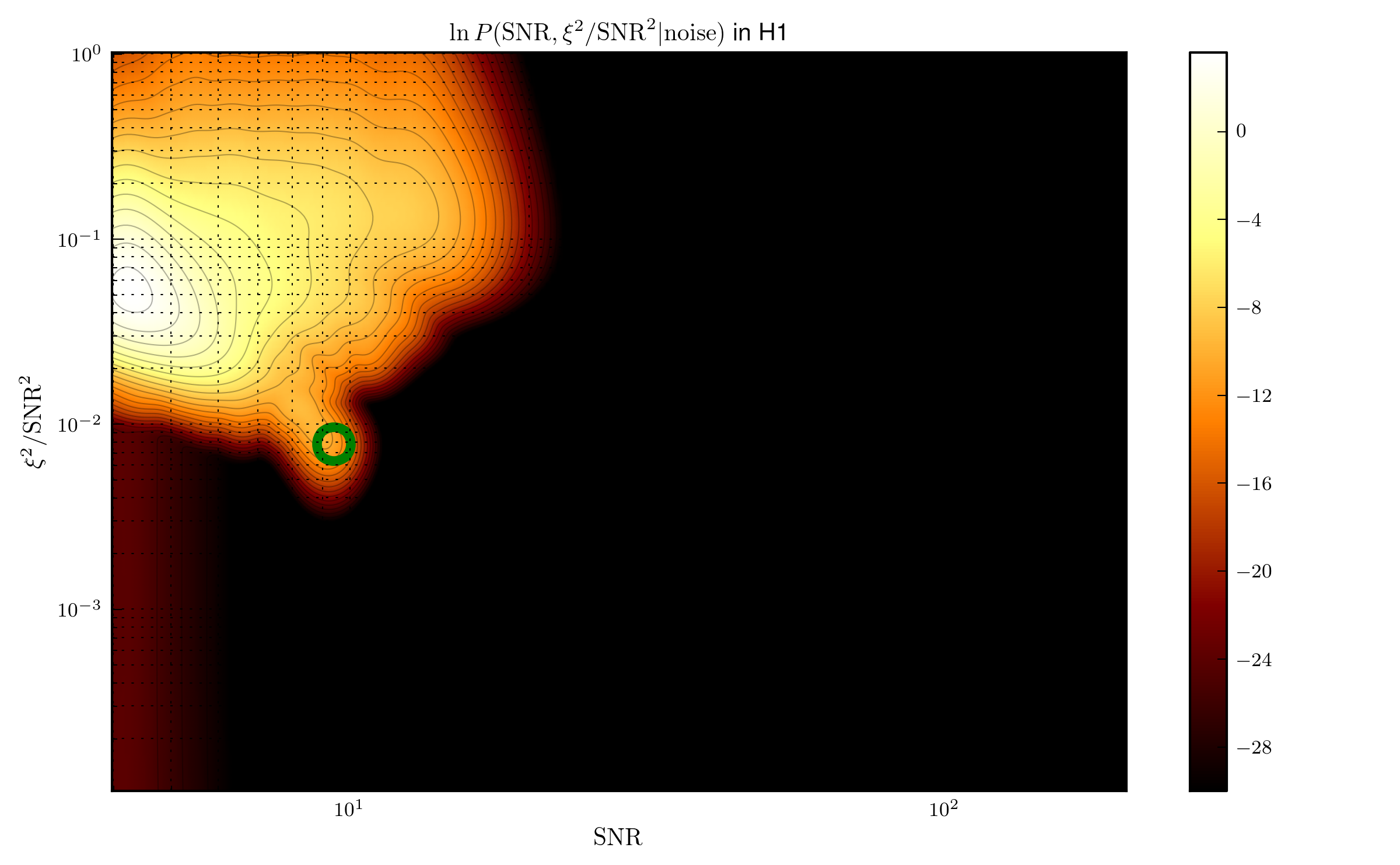}}
	\includegraphics[width=1.0\linewidth]{{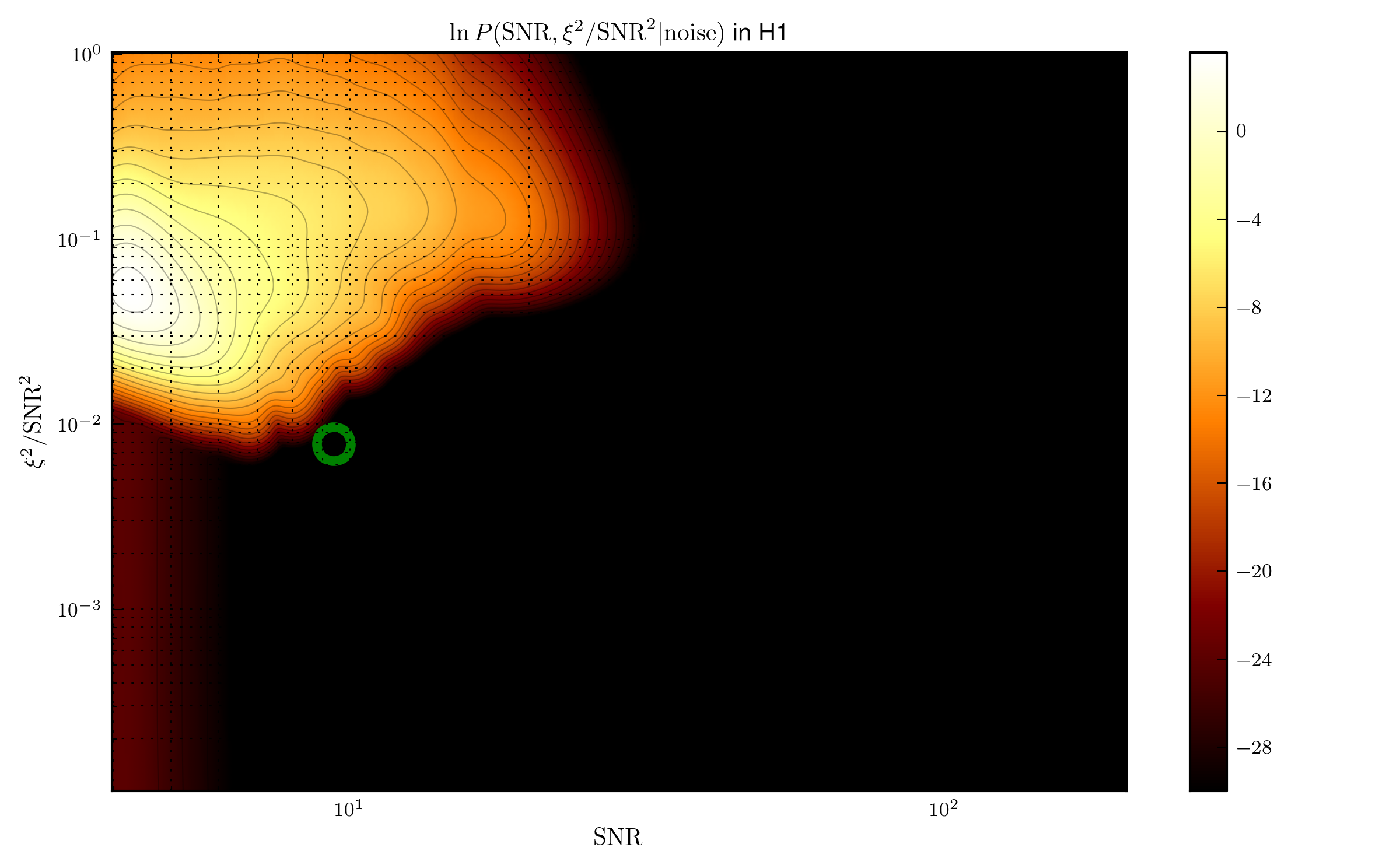}}
	\caption{The background PDFs for the same template bank bin using different
	techniques. The top and bottom figures are the background PDFs obtained from
	the data with four injections using the original tehcnique and improved
	technique respectively. There is a small island, which is the contamination
	due the signal-like triggers, on the top figure, but it is not found on the
	bottom figure; the green circle is where the location of the identified event
	on the PDF.}
	\label{fig:signal_contamination_PDFs}
\end{figure}

The reason for the original search not being able to identify the event is that
the \ac{SNR} and $\xi^2$ values due to that injection were considered as the
background samples. This can be seen from the corresponding background ($\rho,
\xi^2/\rho^2$) PDFs: Figure \ref{fig:signal_contamination_PDFs}. There is an
``island'' on the top figure (contaminated background PDF) after the 4
injections were added to the data, whereas there was no ``island'' on the
bottom figure (recoverd background PDF).

However, the PDF of the recovered background is extended along \ac{SNR} axis,
suggesting that the statistic of \ac{SNR} is not properly estimated by the improved
method. Nevertheless, for this analysis, it did not show any sign of negative
impacts. Finding out the whether the difference will cause any problem for
calculating the likelihood ratio, FAP and/or FAR will be left as future works,
since the time-reversed template bank has shown its robustness to avoid
contamination due to signals.

\begin{figure}
	\centering
	\includegraphics[width=\linewidth]{{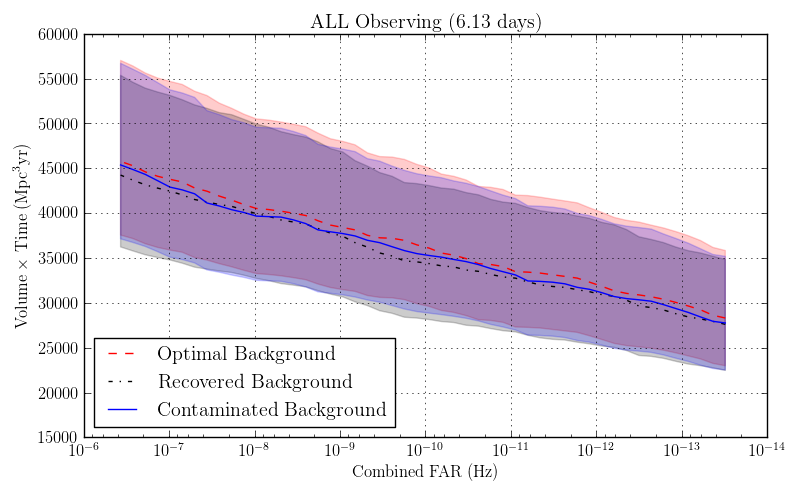}}
	\caption{The plot of the sensitive volume-time versus the combined FAR for
	the three different backgrounds. They are all consistent with each other,
	suggesting that they are all achieving optimal pipeline sensitivity,
	suggesting that there are no immediate improvement to the sensitivity in normal
	signal rate.}
	\label{fig:4_VT}
\end{figure}

The plot of sensitivity is shown on Figure \ref{fig:4_VT}. From the figure,
there are no significant improvement on the sensitivity and the results from
different backgrounds are consistent with each other. This implies that small
number of signals cannot deteriorate the sensitivity of the search pipeline
(unlike the injection set A). At the present rate of significant triggers
observed by GstLAL pipeline (several triggers are recorded into GraceDB with
FAR $\leq 3.85 \times 10^{-7}$ in a week) \cite{gracedb}, the improved
background estimation technique might not show immediate improvement to the
sensitivity. However, if the background is thought to be contaminated by
signals, the improved technique can be used to reveal the uncontaminated
background model and assign correct likelihood ratio and FAR.

%%%%%%%%%%%%%%%%%%%%%%%%%%%%%%%%%%%%%%%%%%%%%%%%%%%%%%%%%%%%%%%%%%%%%%%%%%%%%%%%
\section{Conclusion}
We have shown the use of time-reversed template bank to estimate the background
model for the searches of GWs from binary neutron star coalescence. We
demonstated the improve method with an injection analysis and showed that it can
estimate the background model as if it was estimated on the strain data without
any signal in presence. However, it is not identically the same, but the search
results did not show any sign that they were negatively affected this
inaccuracy. Lastly, we demonstrated an application of the new method at a
realistic signal rate of one BNS signal per 1.75 days, and showed that it can
be used to reveal signals that are originally hidden due to the contamination
of background model.

%%%%%%%%%%%%%%%%%%%%%%%%%%%%%%%%%%%%%%%%%%%%%%%%%%%%%%%%%%%%%%%%%%%%%%%%%%%%%%%%
\begin{acknowledgments}
The authors are grateful for computational resources provided by the LIGO
Laboratory and supported by National Science Foundation Grants PHY-0757058 and
PHY-0823459. This research has made use of data, software and/or web tools
obtained from the Gravitational Wave Open Science Center
(https://www.gw-openscience.org), a service of LIGO Laboratory, the LIGO
Scientific Collaboration and the Virgo Collaboration. LIGO is funded by the
U.S. National Science Foundation. Virgo is funded by the French Centre National
de Recherche Scientifique (CNRS), the Italian Istituto Nazionale della Fisica
Nucleare (INFN) and the Dutch Nikhef, with contributions by Polish and
Hungarian institutes.
\end{acknowledgments}

\bibliography{references}% Produces the bibliography via BibTeX.

\end{document}

%% file: acros.tex
\acrodef{BNS}{binary neutron star}
\acrodef{CBC}{compact binary coalescence}
\acrodef{FAR}{false alarm rate}
\acrodef{FAP}{false alarm probability}
\acrodef{GW}{gravitational-wave}
\acrodef{LVC}{LIGO Scientific and Virgo Collaborations}
\acrodef{MBTA}{Multi-Band Template Analysis}
\acrodef{SNR}{signal-to-noise ratio}